\documentclass[showpacs,showkeys,amsmath,amssymb,aps,twocolumn,pre]{revtex4-1}
\usepackage{graphicx}
\usepackage{dcolumn}
\usepackage{bm}
\begin{document}

\preprint{}

\title{Life and death of stationary linear response in anomalous continuous time random
walk dynamics }

\author{Igor Goychuk}
\thanks{Supported by the Deutsche Forschungsgemeinschaft (German Research Foundation), 
Grant GO 2052/1-2}

 \email{e-mail: igoychuk@uni-potsdam.de}
 
\affiliation{Institute for Physics and Astronomy, University of Potsdam, 
Karl-Liebknecht-Str. 24/25,
14476 Potsdam-Golm, Germany}

\date{\today}

\begin{abstract}
Linear theory of stationary response in systems at thermal equilibrium requires to find
equilibrium correlation function of  unperturbed responding system. 
Studies of the response of the systems exhibiting anomalously slow dynamics 
are often based on the continuous time random walk description (CTRW) with divergent mean
waiting times. The bulk of the literature on anomalous response contains 
linear response functions like one by Cole-Cole calculated from such a CTRW theory and applied
to systems at thermal equilibrium.
Here we show within a fairly simple and general model that for the systems  with divergent 
mean waiting times the stationary response at thermal equilibrium is absent, 
in accordance with some recent studies.
The absence of such stationary response (or dying to zero non-stationary response
in aging experiments) would confirm CTRW with divergent mean waiting times
as underlying physical relaxation mechanism, but reject it otherwise. We show that the
absence of stationary response is closely related to the breaking of ergodicity of the
corresponding dynamical variable. As an important new result, we derive a generalized 
Cole-Cole response within
ergodic CTRW dynamics with finite waiting time. Moreover, we provide a physically
reasonable explanation of the origin and wide presence of $1/f$ noise in condensed matter
for ergodic dynamics close to normal, rather than strongly deviating.
\end{abstract}
\maketitle

\section{Introduction}

Theories of anomalous diffusion and transport abound \cite{KLM}, including continuous time
random walks (CTRW)\cite{Shlesinger,Scher,Hughes} and related 
fractional Fokker-Planck equations \cite{Metzler}, generalized Langevin 
equations \cite{Kubo,Coffey,WeissBook,Goychuk12}, anomalous kinetic equations 
of the (linear) Boltzmann equation type \cite{Balescu,BarkaiSilbey}, hopping
and continuous dynamics in disordered and fractal media \cite{Bouchaud,Havlin}, etc. 
Often it is not clear from the
experimental results which theory works better for the case of study. In particular,
theory of anomalous response is often based on CTRW with divergent mean residence
times (MRTs). Often it is seen as a proper explanation of the physical origin of all major
anomalous response functions, including Cole-Cole response \cite{Cole} featuring the so-called
fast or $\beta-$relaxation in complex liquids and glasses. The latter one is relatively 
fast (lasting 
from pico- to milliseconds, even if it follows to a power law), 
and the corresponding response quickly
becomes stationary on the time scale of observations. It is followed by $\alpha-$relaxation,
which is often described by Kohlrausch-Williams-Watts stretched exponential dependence,
or even better \cite{Lunken1} by the Cole-Davidson relaxation law. 
$\alpha-$relaxation extends up to $10^4-10^6$ seconds
in a typical aging experiment \cite{Lunken1,Oukris}, until a stationary
response regime is reached \cite{Lunken1,Oukris,Lunken}.

Stationary response requires to calculate stationary autocorrelation function (ACF).
If to do this properly for CTRW with divergent mean residence times, it becomes clear 
that such a response is absent, because the corresponding stationary ACF
does not decay at all. This in turn indicates the breaking of ergodicity in accordance 
with the Slutsky theorem \cite{Papoulis}.  
Non-ergodic CTRW cannot respond to stationary perturbations
in a very long run. Its response must die to zero, which invalidates a part of the 
CTRW based theories of anomalous stationary response, when applied to physical
systems at thermal equilibrium. We discuss these subtleties below within a sufficiently
simple and generic model.

\section{Theory and Model}

Linear response theory has been put by Kubo on the firm statistic-mechanical foundation 
\cite{Kubo,Zwanzig,Marconi}. 
Consider a physical system with dynamical variable $x(t)$, say coordinate of a particle
moving in a multistable potential in a dissipative environment.  It exhibits  stochastic
dynamics (here classical) due to thermal agitations of the medium at temperature $T$, with the ensemble 
average $\langle x(t)\rangle_{\rm st}=x_{\rm eq}$ (we bear in mind 
many independent particles). 
The system is at thermal equilibrium. One applies a 
periodic driving force $f(t)=f_0\cos(\Omega t) $ with frequency $\Omega$ and amplitude 
$f_0$ linearly coupled to the dynamical variable $x(t)$, with perturbation energy $-fx$. 
The mean deviation from equilibrium, 
$\langle \delta x(t)\rangle=\langle x(t)\rangle -x_{\rm eq} $, responds generally to the
driving switched on at the time $t_0=0$ with some delay \cite{Kubo,Zwanzig},
\begin{eqnarray}
\langle \delta x(t)\rangle=\int_{0}^t \chi(t-t')f(t')dt',
\end{eqnarray}
where $\chi(t-t')$ is the linear response function (LRF), and the upper integration limit $t$
reflects causality. For any physical system at thermal equilibrium and weakly perturbed
by driving, LRF can depend only on the difference of time arguments. Stationary response
to the periodic signal reads ($t\to\infty$)
\begin{eqnarray}
\langle \delta x(t)\rangle_{\rm as}=f_0|\hat \chi(\Omega)|\cos(\Omega t-\varphi),
\end{eqnarray}
where $\hat \chi(\omega)=\int_{-\infty}^\infty e^{i\omega t}\chi(t)dt=
\int_{0}^\infty e^{i\omega t}\chi(t)dt$ is the 
(Laplace-)Fourier transform of $\chi(t)$ ($\chi(t)=0$ for $t<0$ due to causality), 
and $ \varphi=\tan^{-1}\left({\rm Im}\hat \chi(\omega)/{\rm Re}\hat \chi(\omega)\right )$ 
is the phase lag.
Kubo derived from microscopic
Hamiltonian dynamics under fairly general conditions that such stationary LRF is related
to equilibrium autocorrelation function, or rather covariance $K(t)=
\langle \delta x(t) \delta x(0)\rangle_{\rm eq}$ of the variable $x(t)$ as 
\cite{Kubo,Zwanzig,Marconi}
\begin{eqnarray}\label{FDTtime}
\chi(t)=-\frac{H(t)}{k_BT}\frac{d}{dt}\langle \delta x(t) \delta x(0)\rangle_{\rm eq}\;,
\end{eqnarray}
where $H(t)$ is the Heaviside step function.
This fundamental result in statistical physics
is known as classical fluctuation-dissipation theorem (FDT)
in the time domain.
Stationary response is related to
the equilibrium autocorrelation function which must be stationary. A generalization
of this result to a nonstationary, e.g. aging environment, or in the presence of
steady state nonequilibrium fluxes in the unperturbed state (non-equilibrium steady
state -- NESS) is not trivial.  
Especially, the systems exhibiting non-Gaussian anomalous diffusion demonstrate profound
deviations from  equilibrium FDT behavior in NESS \cite{Villamaina,Gradenigo}.
Non-equilibrium FDT was developed primarily for the
systems with underlying Markovian dynamics \cite{Cugliandolo,Lippielo}. 
It allows, for example, to rationalize 
the NESS response of anomalous 1d subdiffusive dynamics to a small additional
tilt on a comb lattice \cite{Villamaina}. This dynamics
is embedded as 2d Markovian dynamics \cite{Villamaina}.
A generalization to intrinsically non-Markovian dynamics is yet to be done.
In the present work, we are focusing on a standard response
of unperturbed  systems being at thermal equilibrium, 
without dissipative fluxes present.

Specifically, our focus is on anomalous non-Debye type response featuring many
materials. As an important example, famous Cole-Cole  response function \cite{Cole}
reads (setting the high-frequency component $\hat \chi(\infty)$ to zero) 
\begin{eqnarray}
\hat \chi(\omega)=\frac{\chi_0}{1+(-i\omega\tau_r)^\alpha}
\end{eqnarray}
in the frequency domain, 
with static $\chi_0=\langle \delta x^2(0)\rangle_{\rm eq}/(k_BT)$ and $0<\alpha<1$. 
The Cole-Cole response corresponds to the Mittag-Leffler relaxation \cite{Metzler,Gorenflo} 
of the autocorrelation function
\begin{eqnarray}
K(t)=\langle \delta x^2(0)\rangle_{\rm eq} E_{\alpha}[-(t/\tau_r)^\alpha],
\end{eqnarray}
where $E_{\alpha}[z]=\sum_{n=0}^{\infty}z^n/\Gamma(\alpha n+1)$ is Mittag-Leffler function 
\cite{Metzler,Gorenflo}
and $\tau_r$ is (anomalous) relaxation time, which  in the range of from picoseconds 
to milliseconds for various materials \cite{Cole}. 
If a constant signal $f_0=const$ is switched on at $t_0=0$, then 
the time-dependent response in accordance with FDT is
\begin{eqnarray}\label{static}
\langle \delta x(t)\rangle=\frac{f_0}{k_BT}\left [ K(0)-K(t)\right ]\;.
\end{eqnarray}
It starts from zero and approaches the asymptotical value 
$\langle \delta x(\infty)\rangle=\frac{f_0}{k_BT} \langle \delta x^2(0)\rangle_{\rm eq}$, 
if $K(t)\to 0$ with $t\to\infty$.

\subsection{CTRW model}

We model further the thermal dynamics of $x(t)$ by a non-Markovian (semi-Markovian) 
CTRW of the following type. 
There are $N$ localized
states $x_i$ which are populated with stationary probabilities $p_i$. Scattering events
due to a thermal agitation are
not correlated and distributed in time with the waiting time density (WTD)  $\psi(\tau)$, 
which is normalized
and has a finite mean time $\langle \tau\rangle$ between two subsequent events.
Scattering can lead to any other state, but the particle can remain also in the
same state. 
The finiteness of  $\langle \tau\rangle$
is crucial for calculating the stationary ACF.
Then, we can study the response of our system in the limit $\langle \tau\rangle\to \infty$,
\textit{after} the stationary ACF $K(t)$ is found and not \textit{before}, overcoming thereby the common
fallacy done in huge many previous works (they are simply not based on the correct 
stationary ACF). The transition probability from the state $j$ to the state $i$
is $T_{ij}=p_i$, i.e. it corresponds to the statistical weight of the next trapping
state. Such a CTRW
is a particular version of separable CTRW by Montroll and Weiss \cite{Hughes}. 
This is a toy model, but the main features are expected to be generic.  To calculate 
the stationary and equilibrium averages the distribution of the first residence time is required 
\cite{Tunaley,GH03,GH05}. It is well-known to be \cite{Cox}
$\psi^{(0)}(\tau)=\Phi(\tau)/\langle \tau\rangle$ (see also below in Sec. \ref{aging_part}
for a derivation of this result), where $\Phi(\tau)=\int_\tau^\infty
\psi(\tau')d\tau'$ is survival probability. The survival probability for the
first sojourn without scattering events is $\Phi^{(0)}(\tau)=\int_\tau^\infty
\psi^{(0)}(\tau')d\tau'=\int_\tau^\infty
\Phi(\tau')d\tau'/\langle \tau\rangle$. It plays the central role.  We show in 
Appendix that
\begin{equation}\label{ACF1}
K(t)=\langle \delta x^2(0)\rangle_{\rm eq} \Phi^{(0)}(t)
\end{equation}
within this fully decoupled CTRW model.

Now we can ask the question if such a probability density 
$\psi(\tau)$ exists within this model that we can obtain the Cole-Cole response exactly.
The answer is no. Indeed, the Laplace transform of $\Phi^{(0)}(t)$ is related to the Laplace 
 transform of survival probability $\tilde\Phi(s) $ as
\begin{equation}\label{rel1}
\tilde \Phi^{(0)}(s)=\frac{1}{s} \left [ 1- \frac{\Phi(s)}{\langle \tau \rangle}\right ]\;.
\end{equation} 
For the Mittag-Leffler relaxation $\tilde \Phi^{(0)}(s)=\tau_r/[s\tau_r+(s\tau_r)^{1-\alpha}]$.
From this and (\ref{rel1}), $\tilde \Phi(s)=\tau_r/(1+(s\tau_r)^{\alpha})$, which does not
correspond to any survival probability since $\lim_{t\to 0} \Phi(t)=\lim_{s\to \infty} s\tilde 
\Phi(s)=\infty$, and not $\Phi(0)=1$, as must be. Remarkably, 
within the framework of Langevin  dynamics with memory
the Cole-Cole
response is easily reproduced by fractional overdamped Langevin dynamics in parabolic potential
\cite{Goychuk07}.

From Eqs. (\ref{ACF1}) and (\ref{rel1}) it follows immediately in the formal limit 
$\langle \tau \rangle \to\infty$ that
$\tilde K(s)\to \langle \delta x^2(0)\rangle_{\rm eq}/s$  for any WTD, which means that
\begin{equation}\label{ACF-ML}
K(t)=\langle \delta x^2(0)\rangle_{\rm eq}\;.
\end{equation} 
It does not decay at all! From this in (\ref{FDTtime}) it follows immediately that stationary response
is absent exactly,
\begin{eqnarray}
\chi(t)=0 \;.
\end{eqnarray}
 The ``no stationary response'' theorem for a rather general class of processes 
with infinite mean residence time 
is thus proven. It agrees with the death of response found earlier for other 
non-ergodic CTRW dynamics \cite{Barbi,Sokolov06,H07,West,Allegrini09}.
In fact, it follows at once by the fluctuation-dissipation theorem 
(\ref{FDTtime}) from nonergodicity of any process $x(t)$ with divergent
mean residence times \cite{BarkaiMargolin}. The breaking of ergodicity in turn 
follows by the virtue of Slutsky theorem  from nondecaying character of 
stationary autocorrelation function \cite{Papoulis}. 
Non-ergodic system with $\langle \tau\rangle=\infty$ cannot respond stationary to
a stationary (e.g. periodic) signal,
and the corresponding theories of linear response at thermal equilibrium 
are deeply flawed. However,
it is more instructive to investigate the way how the stationary response becomes increasingly 
suppressed with increasing $\langle \tau\rangle$. Two particular models are interesting in this respect.

\subsection{Tempered distribution}

The first model corresponds to the Mittag-Leffler density tempered by introduction of an exponential
cutoff \cite{Saichev}.  The corresponding survival probability is
$\Phi(t)=E_\alpha[-r_\alpha t^\alpha]\exp(-r t)$, where $r_\alpha$ is fractional
rate and $r$ is a cutoff rate. All the moments then become finite.  The survival 
probability has Laplace transform
\begin{eqnarray}
\tilde \Phi(s)=\frac{1}{s+r+r_\alpha(s+r)^{1-\alpha}}\,.
\end{eqnarray}
The mean time is $\tilde \Phi(0)=1/[r+r_\alpha r^{1-\alpha}]=\langle \tau \rangle$, and
\begin{eqnarray}
\tilde K(s)=\frac{\langle \delta x^2(0)\rangle_{\rm eq}}{s}
\frac{s+r_\alpha(s+r)^{1-\alpha}-r_\alpha r^{1-\alpha}}{s+r+r_\alpha(s+r)^{1-\alpha}}\,.
\end{eqnarray}
Clearly, it also exhibits the generic death of stationary response in the limit 
$\langle \tau \rangle\to \infty$.

We shall not elaborated this model in more detail here, bur rather focus on a
very different model with finite $\langle \tau\rangle$, but divergent higher moments.
It exhibits a number of other surprising and interesting features.

\subsection{Mixing normal and anomalous relaxation }

We assume that transition events  can occur  either
with normal rate $r$, or with some fractional rate $r_{\alpha_i}$ ($0<\alpha_i<1$) 
intermittently through $m+1$ independent relaxation 
channels. This yields the following survival probability \cite{Goychuk12PRE},
\begin{eqnarray}
\tilde \Phi(s)=\frac{1}{s+r+\sum_{j=1}^m r_{\alpha_j} s^{1-\alpha_j}}\;,
\end{eqnarray}
which can also be written as
\begin{eqnarray}
\tilde \Phi(s)=\frac{\langle \tau\rangle }{1+\langle \tau\rangle s+
\sum_{j=1}^m (\tau_j s)^{1-\alpha_j}}\;,
\end{eqnarray}
where $\langle \tau \rangle =\tilde \Phi(0)=1/r$ is the mean relaxation 
time  (defined exclusively
by the normal relaxation channel) and the
time constants $\tau_i=(r_{\alpha_i} \langle \tau \rangle)^{1/(1-\alpha_i)}$ 
provide the spectrum of anomalous relaxation times. Specifying the generalized master equation
(GME) by Kenkre, Montroll, and Shlesinger  \cite{Kenkre} for this model (see e.g. Appendix
in Ref. \cite{Goychuk04} for a general derivation from the trajectory perspective) 
yields fractional relaxation equation
\begin{eqnarray} \label{FME}
\dot p_i(t) & = &-\sum_{j=0}^m r_{\alpha_j}\sideset{_0}{_t}{\mathop{\hat
D}^{1-\alpha_j}} \left [ p_i(t)- p_i \right ],
\end{eqnarray}
where
\begin{eqnarray}\label{RL} 
\sideset{_{t_0}}{_t}{\mathop{\hat D}^{1-\alpha}}p(t):=\frac{1}{\Gamma(\alpha)}
\frac{d}{dt }\int_{t_0}^t dt' \frac{p(t')}{(t-t')^{1-\alpha}},
\end{eqnarray}
defines the operator of fractional Riemann-Liouville derivative of the order $1-\alpha$ \cite{Metzler,Gorenflo}.
Eq. (\ref{FME}) belongs to a general class of GMEs 
with distributed fractional derivatives \cite{distributed}. 
One relaxation channel must be normal, $\alpha_0=1$, $r_{\alpha_0}=r$ 
for ergodic process. Evolution of state probabilities $p_i(t)$ become completely decoupled
because of our choice of the transition probabilities, $T_{ij}=p_i$. From Eq. (\ref{FME}),
 $p_i$ are obviously the stationary
probabilities, $\lim_{t\to\infty}p_i(t)=p_i^{(\rm st)}=p_i$, which is not immediately
obvious from the trajectory description. 

For normal relaxation, our choice of model corresponds
to a very popular textbook model of single relaxation time $\langle \tau\rangle$, which does not
depend on the particular state. It is often used as the 
simplest approximation in studying various kinetic equations. By the same token, anomalous relaxation times 
also do not depend on the particular state within the model considered.
Notice, however, that GME (\ref{FME}) cannot be used to find the stationary ACF (\ref{ACF1}).
Its direct use would be namely the typical fallacy which plagued many earlier works.  One
must proceed differently, see in Appendix for details.

\begin{figure}
\includegraphics[width=7.5cm]{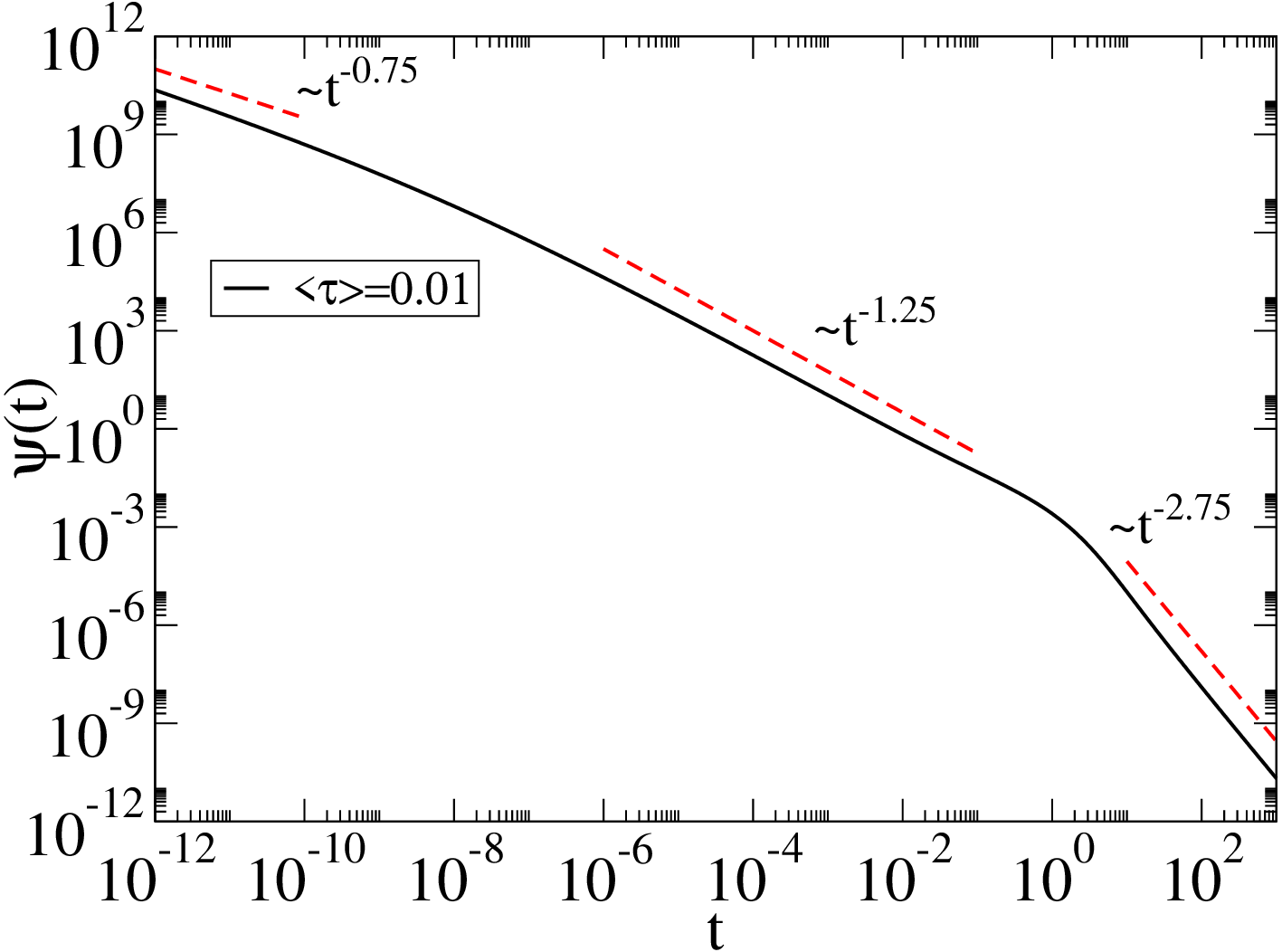}
\caption{ Dependence of probability density $\psi(\tau)$
on time (in units of $\tau_r$) for  $\langle \tau\rangle=0.01$ and $\alpha=0.25$
exhibits three different power laws, $\tau^{-1+\alpha}$, $\tau^{-1-\alpha}$, $\tau^{-3+\alpha}$, 
in the parameter regime $\langle \tau\rangle\ll \tau_r$.
 }
\label{Fig1}
\end{figure}

\begin{figure}
\includegraphics[width=7.5cm]{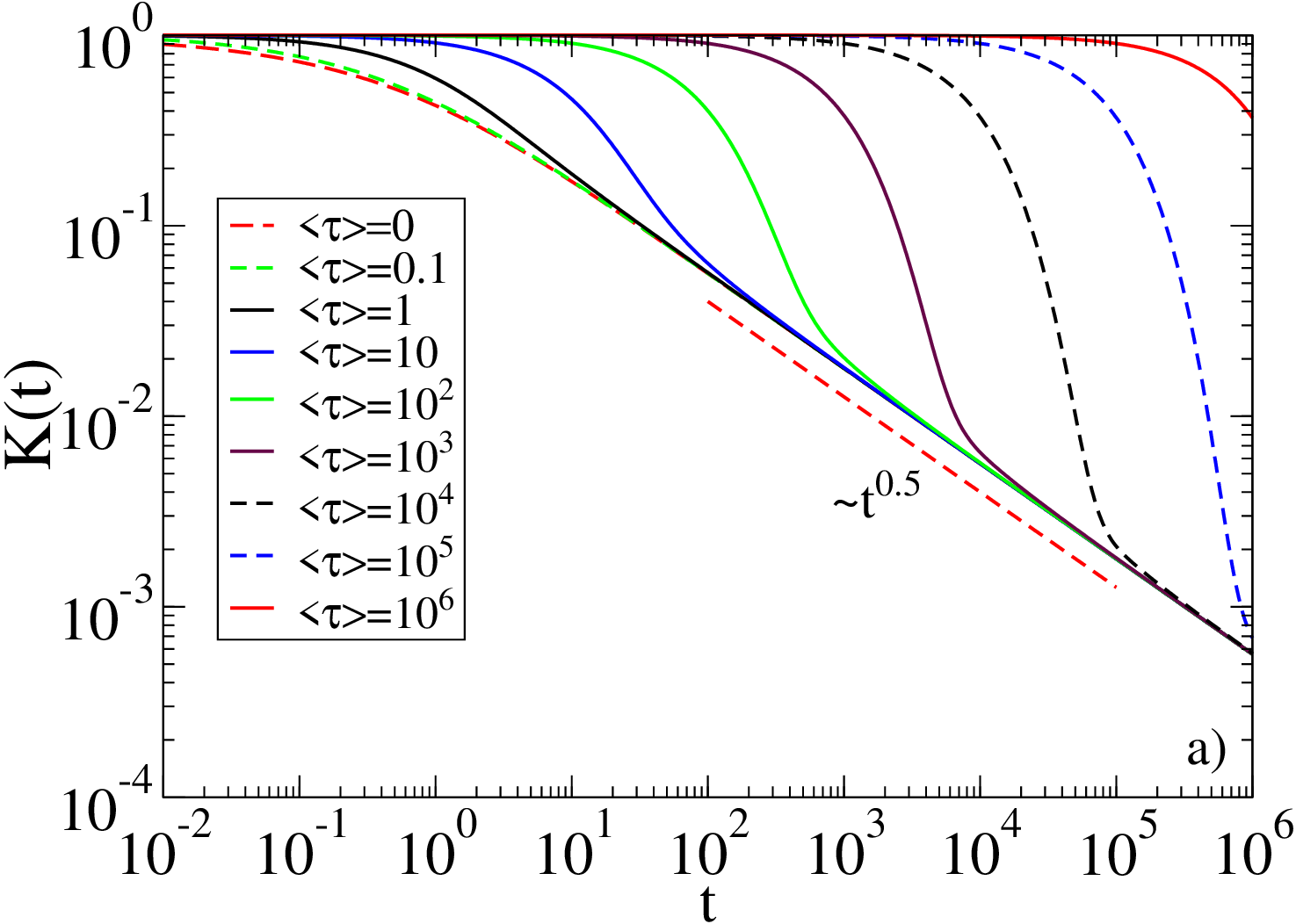}\\ \vspace{1cm}
\vfill
\includegraphics[width=7.5cm]{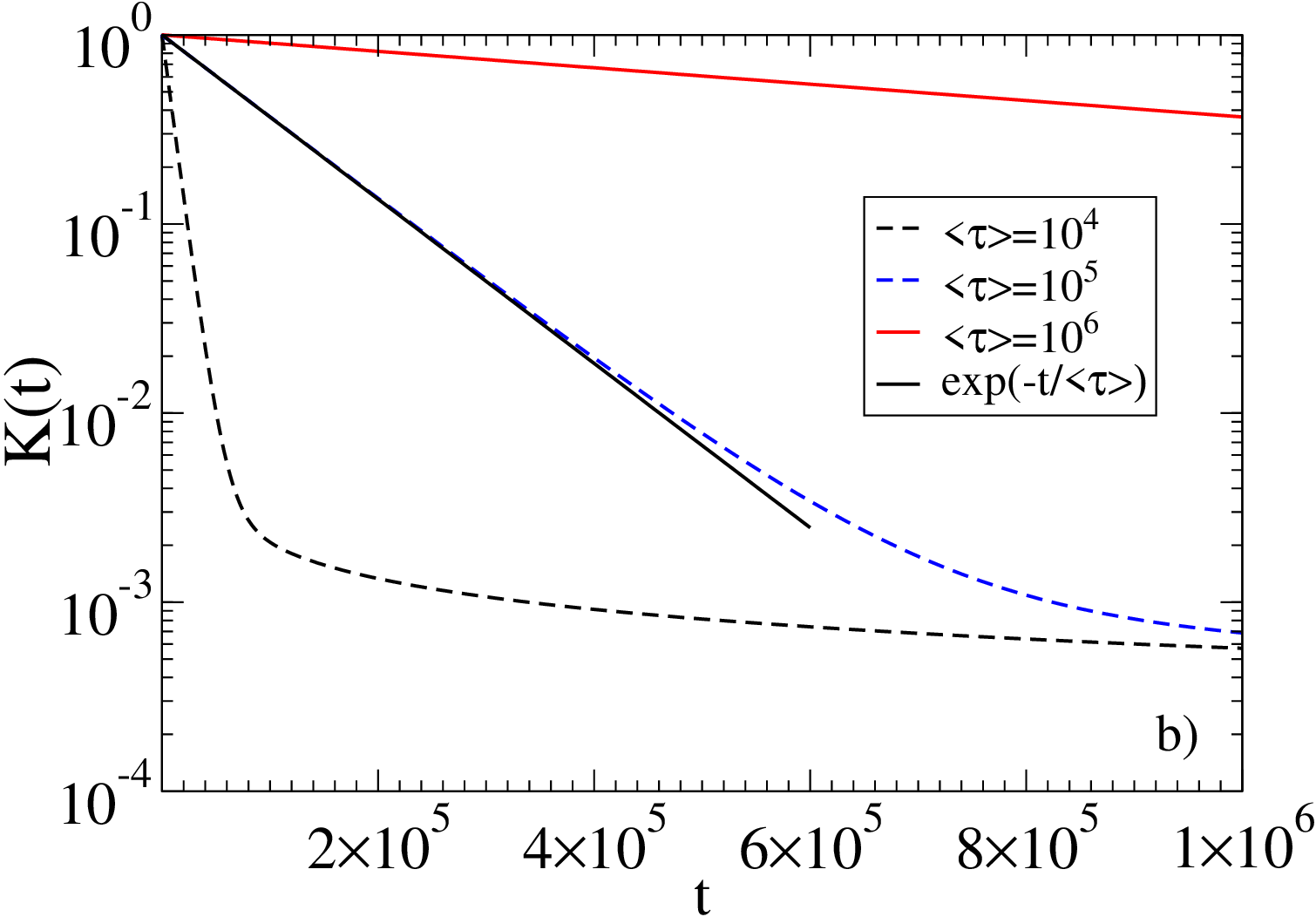}
\caption{ Normalized correlation function (in units of $\langle \delta x^2(0)\rangle_{\rm eq}$)
on time (in units of $\tau_r$) for different $\langle \tau\rangle$ and $\alpha=0.5$.
The tail is always $K(t)\propto t^{-1+\alpha}$, see in part a). Decay of correlations is dramatically delayed
with the increase of $\langle \tau\rangle$. Initially, it is exponential for 
$\langle \tau\rangle \gg \tau_r$, $K(t)\approx \langle \delta x^2(0)\rangle_{\rm eq}
\exp(-t/\langle \tau\rangle)$, as part b) shows. The limit $\langle \tau\rangle\to 0$, corresponds
to the Mittag-Leffler decay of correlations, $K(t)\approx \langle \delta x^2(0)\rangle_{\rm eq}
E_{1-\alpha}[-(t/\tau_r)^{1-\alpha}]$ and Cole-Cole response.  Part b) represents some of
the curves in part a) on semi-logarithmic plot revealing initial exponential decay 
for large $\langle \tau\rangle$.\\
 }
\label{Fig2}
\end{figure}

For the simplest nontrivial case $m=1$ with $\alpha_1=\alpha$,
$\tau_r=\tau_1$, and in the parameter regime $\langle \tau \rangle \ll \tau_r$,
 $\Phi(t)\approx E_\alpha(-r_\alpha t^\alpha)$, 
 on the time scale  $t\ll \tau_r$. It corresponds to the same WTD 
as in the basic CTRW model with divergent $\langle \tau\rangle$. Hence,
for $t\ll \tau_\alpha=r_\alpha^{-1/\alpha}$,  
$\psi(\tau)\propto \tau^{-1+\alpha}$. This initial power law
corresponds to the initially stretched exponential survival probability 
$\Phi(t)\approx \exp(-r_\alpha t^\alpha/\Gamma(1+\alpha))$. Then, 
 intermediate power law follows, $\psi(\tau)\propto \tau^{-1-\alpha}$,
for $\tau$ within the range 
\begin{eqnarray}
\tau_\alpha=[\langle \tau\rangle/\tau_r]^{1/\alpha}
\tau_r \ll\tau\ll \tau_r \;,
\end{eqnarray}
i.e. over $[\tau_r/\langle \tau\rangle]^{1/\alpha}$ intermediate time decades.
For $\tau\gg \tau_r$, it further changes  into the asymptotic power law
$\psi(\tau)\propto \tau^{-3+\alpha}$ \cite{Goychuk12PRE}.  
Interestingly enough, introduction of a finite $\langle \tau\rangle$ results
into a sufficiently strong power law decay for $\tau >\tau_r$. 
The parameter $\tau_r$ plays thus the role of a cutoff time, though the cutoff
character is very different
from the model of exponentially tempered distribution. The latter one has an exponential cutoff
for $t\gg 1/r$.
The important parameter
regime, $\langle \tau \rangle \ll \tau_r$, is depicted in Fig.~\ref{Fig1}. 
 Notice,
however, that for $\langle \tau \rangle \sim \tau_r$, the intermediate power law first disappears,
and for $\langle \tau \rangle \gg \tau_r$ it transforms into intermediate exponential decay
(like one in Fig. \ref{Fig2},b).

The Laplace-transformed stationary ACF of this model reads
\begin{equation}\label{ACF}
\tilde K(s)=\langle \delta x^2(0)\rangle_{\rm eq} \frac{1+ 
\sum_{i=1}^m r_{\alpha_i} s^{-\alpha_i}}{s+r + \sum_{i=1}^m r_{\alpha_i} s^{1-\alpha_i}},
\end{equation}
from which it becomes immediately clear that in the limit $r=1/\langle \tau \rangle\to 0$, 
the limiting stationary ACF is simply constant, as in Eq. (\ref{ACF-ML}). The stationary
response is absent.

Furthermore, the response function in Laplace domain follows as
\begin{equation}\label{response}
\tilde \chi(s)=\frac{\chi_0 }{1+\langle \tau \rangle s +\sum_{i=1}^m( \tau_i s)^{1-\alpha_i}}.
\end{equation}
 In frequency domain it is 
$\hat \chi(\omega)=\tilde \chi(-i\omega)$. We have thus a nice and nontrivial generalization
of Cole-Cole response function which includes one normal relaxation time 
$\langle \tau \rangle$ and $m$ anomalous. It provides a very rich model of anomalous
response based on ergodic CTRW dynamics with finite mean waiting time. 
We consider further the particular case of one anomalous relaxation channel, $m=1$.
 Then,
the Cole-Cole response with index $1-\alpha$ instead of $\alpha$ is reproduced 
in the limit $\langle \tau \rangle\to 0$ at fixed anomalous relaxation time $\tau_r=\tau_1$. 
Strikingly enough, this is the opposite limit with respect to one 
considered in the theory of CTRW with infinite $\langle \tau \rangle=\infty$. Indeed,
for sufficiently small $\langle \tau\rangle\ll \tau_r$,
$K(t)\approx \langle \delta x^2(0)\rangle_{\rm eq} E_{1-\alpha}[-(t/\tau_r)^{1-\alpha}]$,
see in Fig. \ref{Fig2}. This corresponds to the Cole-Cole response with exponent $1-\alpha$
providing a very important result: stationary Cole-Cole response emerges within 
the CTRW approach if there exists a very
fast normal relaxation channel acting in parallel to the anomalous one and making 
the mean transition
times finite. In striking contrast to the traditional 
CTRW models based on $\psi(\tau)\propto \tau^{-1-\alpha}$ dependence 
with divergent mean residence times our approach yields the Cole-Cole response with exponent 
$1-\alpha$ instead of $\alpha$.

\begin{figure}
\includegraphics[width=7.5cm]{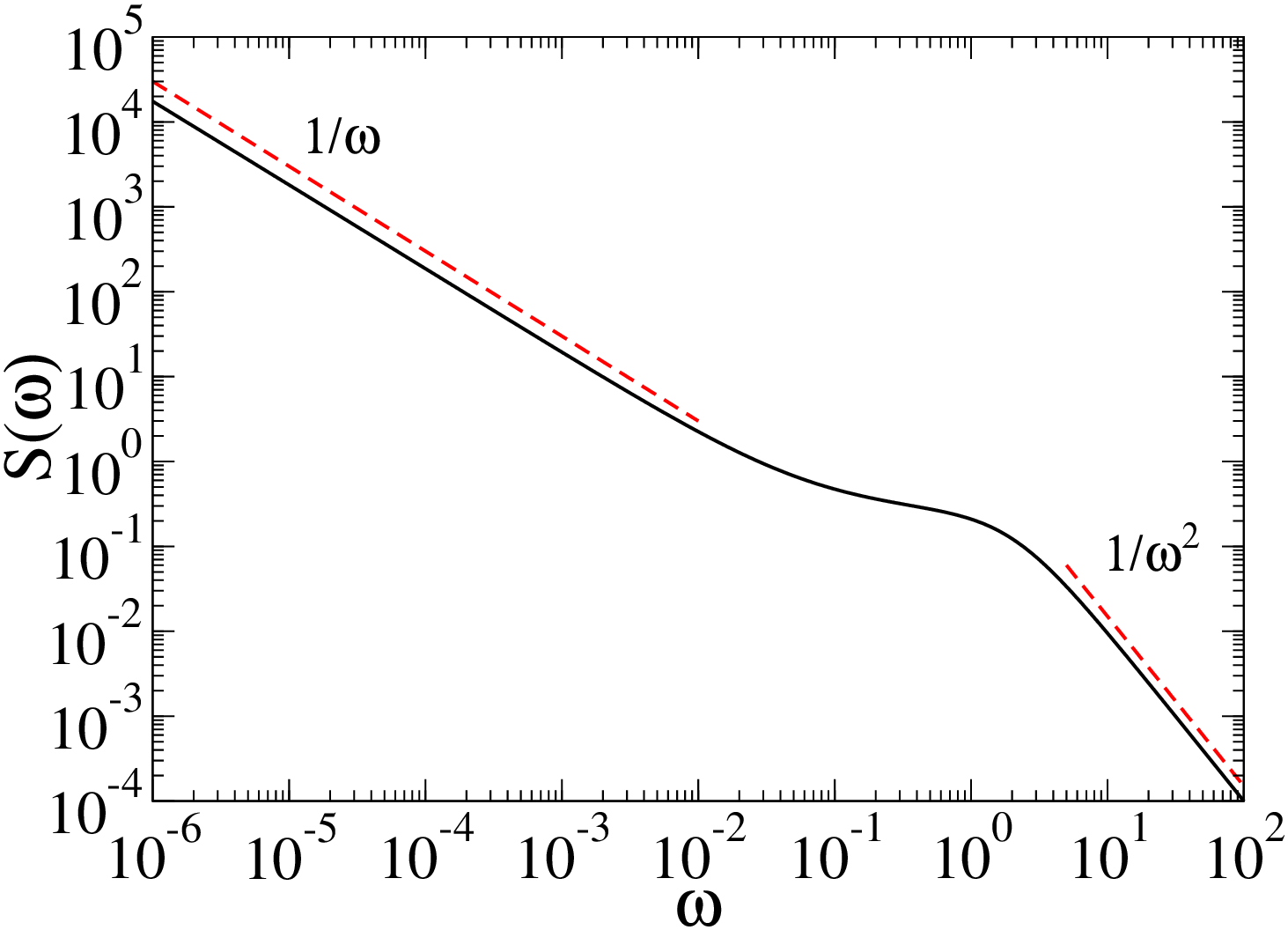}
\caption{ Power spectrum $S(\omega)$ (in units of $2\langle \delta x^2(0)\rangle_{\rm eq}\tau_r$)
on circular frequency (in units of $\tau_r^{-1}$) for  $\alpha=0.95$ and $\langle \tau\rangle=\tau_r$.
 }
\label{Fig3}
\end{figure}

Furthermore, it is instructive to see how $K(t)$ and the 
response function $\hat \chi(\omega)$, behave for large but finite $\langle \tau \rangle =1/r$ 
with $\tau_r$ kept constant. 
Autocorrelation function is plotted in Fig. \ref{Fig2} for $\alpha=0.5$. Notice that it has 
a heavy power law tail $K(t)\propto 1/t^{1-\alpha}$, onset of which moves to larger $t$
with the increase of $\langle \tau \rangle$. For very large $\langle \tau \rangle\gg \tau_r$,
the initial decay is nearly exponential, $K(t)\approx \langle \delta x^2(0)\rangle_{\rm eq}
\exp(-t/\langle \tau\rangle)$. With this in Eq. (\ref{static}), one can clearly see
that on the time scale $t\ll \langle \tau \rangle$, the response of stationary equilibrium 
environment is not realized, being increasingly suppressed with the increase of 
$\langle \tau \rangle$. 
Time-dependent signals with frequencies smaller than a corner
frequency $\omega_c=1/\langle \tau \rangle$ should be regarded as slow. However, the
response to them (even if it does exist asymptotically!) cannot be detected 
in noisy background as we shall clarify soon.
The dependence on $\alpha$ is very
important and remarkable. It is very different from one of CTRW theory with infinite
mean time. Basically, we have $\alpha$ instead of $1-\alpha$.
This means that with $\alpha$ close to one, e.g. $\alpha=0.95$,
the corresponding power noise spectrum 
\begin{eqnarray}
S(\omega)&&=2{\rm Re}[ \tilde K(-i\omega)]\nonumber \\
&&= 2\langle \delta x^2(0)\rangle_{\rm eq}\frac{\tau_r(\omega\tau_r)^\alpha \cos(\frac{\alpha \pi}{2})
+\langle \tau \rangle
 (\omega\tau_r)^{2\alpha}}{A(\omega)},
 \end{eqnarray}
 where
\begin{eqnarray} 
 A(\omega)=(\omega\tau_r)^{2\alpha}+ \left (\frac{\langle \tau \rangle}{\tau_r}
 \right )^2(\omega\tau_r)^{2(1+\alpha)}+(\omega\tau_r)^{2}\nonumber \\+2\frac{\langle \tau \rangle}{\tau_r}
 (\omega\tau_r)^{\alpha +2}\cos\left ( \frac{\alpha \pi}{2}\right )+
 2(\omega\tau_r)^{\alpha +1}\sin\left (\frac{\alpha \pi}{2}\right ), 
\end{eqnarray}
becomes very close to $1/f$ noise at small frequencies, $S(\omega)\propto
1/\omega^\alpha$, see in Fig. \ref{Fig3}. Notice that
the closer the anomalous relaxation  channel to the normal one  the closer
 the power spectrum becomes to one  of $1/f$ noise! This provides a very nice and physically
plausible explanation of the wide presence of $1/f$ noise in 
condensed matter \cite{Weisman}. 
The second relaxation channel
must be most similar to the normal one and not mostly deviating.

\begin{figure}
\includegraphics[width=7.5cm]{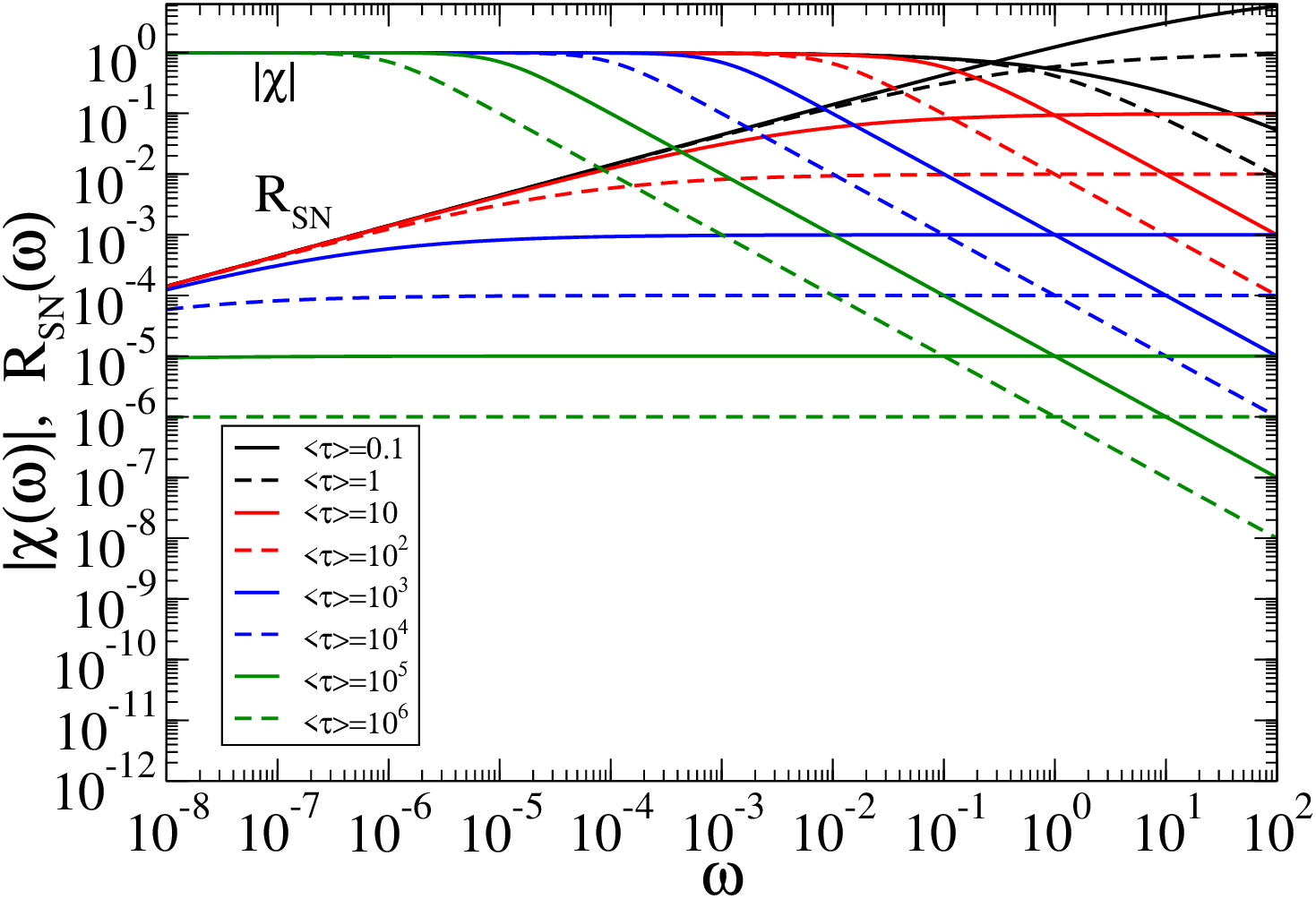}
\caption{ Relative response amplitude  $|\chi(\omega)|$ (in units of $\chi_0$) and signal-to-noise ratio
$R_{\rm SN}(\omega)$ (in units of $B/\tau_r$)
on circular frequency (in units of $\tau_r^{-1}$) for  $\alpha=0.5$ and 
different values of $\langle \tau\rangle$. Notice that both the frequency domain of quasi-static response
and the maximal value of $R_{\rm SN}(\omega)$ shrink as $1/\langle \tau\rangle$ with increase of 
 $\langle \tau\rangle$ (in units of $\tau_r$).
 }
\label{Fig4}
\end{figure}

Furthermore, real and imaginary parts of the response function are
\begin{eqnarray}
{\rm Re }[\chi(\omega)]&&=\chi_0\frac{(\omega\tau_r)^{1+\alpha} \sin(\frac{\alpha \pi}{2})
+(\omega\tau_r)^{2\alpha}}{A(\omega)},
 \end{eqnarray}
and 
\begin{eqnarray}
{\rm Im }[\chi(\omega)]&&=\frac{1}{2k_BT}\omega S(\omega)\label{FDTf}\\
&&=\chi_0\frac{(\omega\tau_r)^{1+\alpha} \cos(\frac{\alpha \pi}{2})
+\frac{\langle \tau \rangle}{\tau_r}(\omega\tau_r)^{2\alpha+1}}{A(\omega)},
 \end{eqnarray}
correspondingly. Notice that Eq. (\ref{FDTf}) is nothing else 
the classical FDT in the frequency domain. This is an exact relation justifying the given
name FDT, since ${\rm Im }[\chi(\omega)]$ is related to dissipation losses. It cannot be violated
at thermal equilibrium for classical dynamics.
The absolute value of the response function measuring the output-to-input ratio
of periodic signal amplitude is 
\begin{eqnarray}
|\chi(\omega) | =\chi_0\frac{(\omega\tau_r)^{\alpha}}{\sqrt{A(\omega)}}, 
 \end{eqnarray}
and the phase lag given by
\begin{eqnarray}
\tan \varphi =\frac{(\omega\tau_r)^{1+\alpha} \cos(\frac{\alpha \pi}{2})
+\frac{\langle \tau \rangle}{\tau_r}(\omega\tau_r)^{2\alpha+1}}{(\omega\tau_r)^{1+\alpha} \sin(\frac{\alpha \pi}{2})
+(\omega\tau_r)^{2\alpha}} \;.
 \end{eqnarray}
 For slow signals with frequencies $\omega\ll \omega_c$, 
 $|\chi(\omega) |\approx \chi_0$. This corresponds to quasi-static response.  
 The response to such slow periodic signals is present, as this result shows. However,
 this does not mean that it can be detected.
The signal-to-noise ratio (SNR), $R_{\rm SN}(\omega)=\pi f_0^2 |\chi(\omega) |^2/S(\omega)$ 
which measures the ratio of spectral amplitude of signal
to the spectral power of noise at the same frequency is important to determine
whether the signal can in principle be detected in the noise background, or not \cite{SR}. 
For the discussed 
model, 
\begin{eqnarray}
R_{\rm SN}(\omega)&&=B \frac{(\omega\tau_r)^{2\alpha}}{\tau_r(\omega\tau_r)^\alpha \cos(\frac{\alpha \pi}{2})
+\langle \tau \rangle (\omega\tau_r)^{2\alpha}}, 
 \end{eqnarray}
where $B=(\pi/2) f_0^2\langle \delta x^2(0)\rangle_{\rm eq}/(k_BT)^2$. For small frequencies SNR
is strongly suppressed since then $R_{\rm SN}(\omega)\propto \omega^\alpha$.
Similar remarkable feature has been detected in the theory of non-Markovian 
Stochastic Resonance \cite{GH03}.
For sufficiently large frequencies, SNR becomes frequency-independent and attains the maximal value 
$R_{\rm SN}^{(\rm max)}(\omega)=B/\langle \tau \rangle$. Clearly, for
 $\langle \tau \rangle\to \infty$, $R_{\rm SN}(\omega)\to 0$, and even slow signals
 cannot be detected in the noise background, cf. Fig. \ref{Fig4}. 
 Besides, the response itself to all signals with frequency $\omega\gg 1/\langle \tau \rangle$
 becomes strongly suppressed. Stationary response is thus virtually 
 absent in  all systems at thermal equilibrium 
 in the limit  $\langle \tau \rangle\to \infty$.

\subsection{Aging correlation function}\label{aging_part}

Nonstationary response exists, however,  even in non-ergodic systems 
\cite{Barbi,Sokolov06,H07,West,Allegrini09}. 
To describe nonstationary response
also in ergodic systems driven initially out of equilibrium and left to relax or age to the new equilibrium
(e.g. after a temperature jump), one needs to
find non-stationary or aging correlation function of the
unperturbed system $\langle x(t_a+t)x(t_a)\rangle$ which depends on the system age $t_a$.
For the considered CTRW model, it can 
also be found straightforwardly following to the derivation given in Appendix. 
One can repeat it with another first-time survival probability 
$\Phi^{(0)}(t|t_a)$ which depends on the age of system $t_a$ instead of 
$\Phi^{(0)}(t)$. This yields
\begin{eqnarray}\label{NACF}
\langle x(t_a+t)x(t_a)\rangle_{0}=\langle \delta x^2 \rangle_{0}
\Phi^{(0)}(t|t_a) + \langle x  \rangle_{0} \langle x \rangle_{\rm st},
\end{eqnarray}
where averaging $\langle ...\rangle_{0}$ is done over initial distribution 
$p_j(0)$ which generally is different from the stationary one, and 
$\langle \delta x^2 \rangle_{0}:=\langle  x^2 \rangle_{0}-
\langle x  \rangle_{0} \langle x \rangle_{\rm st}$. 
The major problem is reduced to finding $\Phi^{(0)}(t|t_a)$.
The central role is played indeed by the first time survival probability, as already well-established
\cite{Godreche,BarkaiMargolin,Allegrini05,West}
for symmetric two state 
process of the kind considered. It features also our fully decoupled 
CTRW model with arbitrary number of states.
Aging $\Phi^{(0)}(t|t_a)$ can be found from the exact relation similar to one
discussed for forward recurrence-time in Refs. \cite{Cox,Allegrini05},
\begin{eqnarray}\label{age1}
\Phi^{(0)}(t|t_a)=&&\Phi(t+t_a) \label{f1}\\
+&&\sum_{n=1}^{\infty}\int_0^{t_a} \psi_{n}(t_a-y)\Phi(t+y)dy \label{sum},
\end{eqnarray}
where $\psi_{n}(t)$ is the probability density to have $n$ scattering events. It is the $n$-time
convolution of the density $\psi(\tau)$, $\tilde \psi_n(s)=[\tilde\psi(s)]^n$ in the Laplace space.
Indeed, let assume that our system was newly prepared at the time $-t_a$ in the past relative
to the starting point of observations $t_0=0$. Then, the first term (\ref{f1}) is just the
probability to stay from $-t_a$ until $t$ without any scattering event. However, $n$ such intermittent  events
can occur until any ``unseen'' time point $-y$ within the time interval $[-t_a,0]$ in the past and then
no events occur until $t$. Integrating over $y$ and summing over all possible $n$ yields
the above exact result. From it, one can easily find the double Laplace transform 
$\tilde  \Phi^{(0)}(s|u)$, where $s$ is Laplace-conjugated to $t$ and $u$ to $t_a$.
Some  algebra yields simple result
\begin{eqnarray}
\tilde  \Phi^{(0)}(s|u)=\frac{1}{u(s-u)}\left( 1-\frac{\tilde\Phi(s)}{\tilde\Phi(u)}\right)\;.
\end{eqnarray}

The Laplace-transform of the fully aged $\Phi^{(0)}(t)=\lim_{t_a\to\infty}\Phi^{(0)}(t|t_a)$, 
the first-time stationary survival probability, 
can be obtained now as  
$\tilde \Phi^{(0)}(s)=\lim_{u\to 0}u \tilde  \Phi^{(0)}(s|u)$. For 
$\tilde\Phi(0)=\langle \tau\rangle\neq \infty$ this reproduces the well-known result in Eq. (\ref{rel1}) 
and provides a very important consistency check. And for $\tilde\Phi(0)=\infty$,
$\tilde \Phi^{(0)}(s)=1/s$, or $\Phi^{(0)}(t|\infty)=\Phi^{(0)}(t)=1$. This entails again
the death of stationary response. In  this case, nonstationary response to a periodic 
signal is dying to zero asymptotically, as found for a two-state nonergodic dynamics 
in Ref. \cite{Barbi}. Ergodic systems with finite $\langle \tau\rangle$ will also exhibit aging response
being prepared in a non-equilibrium state at some $-t_a\neq -\infty$ in the past, e.g. after a temperature
jump. 
Their response but approaches non-zero stationary limiting
 value featuring new
equilibrium state, which has been considered in this work. This is what normally seen
in most aging experiments: response to a periodic signal gradually dies out and approaches a stationary 
non-zero limit \cite{Lunken}.

Nonstationary aging response within the considered model will be studied in detail elsewhere.
As an intelligent guess based on multi-state Markovian dynamics \cite{Marconi,Cugliandolo,Lippielo} in the absence 
of stationary fluxes in unperturbed dynamics, the two-time inhomogeneous response function should read
\begin{eqnarray}\label{FDTneq}
\chi(t,t')=\frac{H(t-t')}{2k_BT}\Big [ \frac{\partial }{\partial t'}\langle \delta x(t) \delta x(t')\rangle \nonumber  \\ 
- \frac{\partial}{\partial t}\langle \delta x(t) \delta x(t')\rangle \Big ]\;,
\end{eqnarray}
with $\langle \delta x(t) \delta x(t')\rangle=\langle \delta x^2 \rangle_{t'}
\Phi^{(0)}(t-t'|t')$, $T$ being the temperature after the temperature quench and 
\begin{eqnarray}
\langle \delta x(t)\rangle=\int_{t_0}^t \chi(t,t')f(t')dt',
\end{eqnarray}
with external field starting to act at $t=t_0$.
As to the survival probability $\Phi(t)$ incorrectly used instead of 
$\Phi^{(0)}(t)$ in most CTRW theories 
of stationary response in thermally equilibrium environments, 
it corresponds to $\Phi^{(0)}(t|t_a)$ at zero age $t_a=0$, 
$\Phi(t)=\Phi^{(0)}(t|0)$, i.e. to the response of systems mostly deviating
from thermal equilibrium,  and not mostly close to it. Clearly, it does not describe even
nonstationary zero-age response in aging systems.

\section{Discussion and Conclusions}

In this work, we showed within a multistate renewal model 
that nonergodic systems featured by infinite mean waiting
times cannot respond asymptotically to stationary signals. This questions a good part
of the anomalous response theory based on such processes. A common fallacy in the corresponding
literature consists in failure
to find the correct stationary autocorrelation function which can be used 
as a phenomenological input in the fundamental microscopical theory of stationary 
linear response by Kubo and others. In this respect, our critique is similar to
one by Tunaley \cite{Tunaley}  earlier.  More important, we studied the way how the
stationary response dies out with increasing mean waiting times in a model 
with one normal and one anomalous relaxation channel acting concurrently and in parallel.
The normal channel defines the mean waiting time.
This model reproduces
approximately the Cole-Cole response in the limit where mean time is much less than
the anomalous relaxation time defining the time of Cole-Cole response (about 
inverse frequency at the maximum of absorption line defined by 
${\rm Im}[\hat \chi(\omega)]\propto \omega S(\omega)$). Paradoxically enough,
such a response is more anomalous, with index $1-\alpha$, for less anomalous relaxation
channels with index $\alpha$ closer to one. Then, the resulting stochastic process
becomes closer to $1/f$ noise. This provides an elegant way to explain
the origin of $1/f$ noise within ergodic dynamics featured by stationary response.

Absence of stationary response for nonergodic dynamics with
infinite mean waiting times does not mean of course that the response is totally absent.
It will be dying down to zero in the course of time as clarified earlier 
for two-state nonergodic dynamics \cite{Barbi,West}. However, it can be of primary importance
for reacting on non-stationary signals only transiently present -- a common situation in many 
biological applications \cite{Goychuk01}, or for other complex nonergodic input signals 
\cite{West}.
For example, response of neuronal systems to boring constant step signals should  normally be damped out
(a healthy reaction),
and this is indeed the case as shown in several experiments \cite{Nature1,French,Nature2}. 
This is a sign of complexity \cite{West}.
A theory of such dying nonstationary response within CTRW approach has been initiated in 
Refs. \cite{Barbi,Allegrini09,West} and we refer interesting readers
to those works. They laid grounds for a continuing  scientific exploration
of the response of non-ergodic, non-stationary and 
aging systems which encompass also this author and readers.  
This fascinating scientific exploration is now at the very beginning. I am confident 
that it will attract ever more attention in the future.

\section*{Acknowledgment} 
 The hospitality of Kavli Institute for Theoretical Physics
at the Chinese Academy of Sciences 
in Beijing, where this work was partially done, is  gratefully acknowledged.

\appendix

\section{Calculation of autocorrelation function}

Stationary function 
$\langle x(t)x(0)\rangle_{\rm st}$ can be found
as $\langle x(t+t_0)x(t_0)\rangle_{\rm st}=\sum_{i,j}x_ix_j P^{(\rm st)}(i,t+t_0;j,t_0)$,
where $P^{(\rm st)}(i,t+t_0;j,t_0)$ is  two-time stationary probability density of the process
$x(t)$ (joint probability). It depends only on the time difference of arguments and can
 be expressed through the 
corresponding stationary conditional probability density or propagator 
$\Pi^{(\rm st)}(i,t|j,0)$ as $P^{(\rm st)}(i,t+t_0;j,t_0)=\Pi^{(\rm st)}(i,t|j,0)p_j^{(\rm st)}(0)$,
where $p_j^{(\rm st)}(0)$ is stationary single-time probability density.
How to construct stationary propagator of such and similar CTRW processes is explained in detail
in Refs. \cite{Goychuk04,GH05}. Along similar lines, we introduce the matrix of waiting time
densities $\mathbf \Psi(\tau)$ which in the present case is expressed through the only one density
$\psi(\tau)$. In components, $\Psi_{ij}(\tau)=\psi(\tau)\delta_{ij}$, where $\delta_{ij}$ is Kronecker
symbol (unity matrix  $\mathbf I$). Next, we introduce the transition matrix $\mathbf T$ with component 
$T_{ij}$ being the transition probability from the state $j$ to the state $i$ as a result of scattering
event.  Matrix of survival probabilities
is $\mathbf \Phi(t)=\Phi(t)\delta_{ij}$. Moreover, we need the matrices of the first time densities,
$\Psi_{ij}^{(0)}(\tau)=\psi^{(0)}(\tau)\delta_{ij}$ 
and the first time survival probabilities, $\Phi_{ij}^{(0)}(t)=\Phi^{(0)}(t)\delta_{ij}$, 
$\Phi^{(0)}(t)=\int_t^{\infty}\psi^{(0)}(t')dt'$.  
Stationary propagator is 
is easy to find in the Laplace-space, $\tilde P^{(\rm st)}(i,s|j,0)=\int_0^{\infty}
P^{(\rm st)}(i,t|j,0)e^{-st}dt$ (tilde denotes the corresponding 
Laplace-transform for any function). Then, the conditional state probabilities to remain in the 
initial state are captured by $\tilde {\mathbf \Phi}^{(0)}(s)$. Contribution of the path with
one scattering event is $\tilde {\mathbf \Phi}(s)\mathbf T \tilde {\mathbf \Psi}^{(0)}(s)$.
Two scattering events contribute as 
$\tilde {\mathbf \Phi}(s)\mathbf T \tilde {\mathbf \Psi}(s)
\mathbf T \tilde {\mathbf \Psi}^{(0)}(s)$, and so on. All in all,
\begin{eqnarray}
\tilde {\mathbf \Pi}^{(\rm st)}(s)=&& \tilde {\mathbf \Phi}^{(0)}(s) \label{first} \\ && +
\tilde {\mathbf \Phi}(s)\sum_{n=0}^\infty\left [\mathbf T \tilde {\mathbf \Psi}(s)\right ]^n
\mathbf T \tilde {\mathbf \Psi}^{(0)}(s) \\
&& =\tilde {\mathbf \Phi}^{(0)}(s)  +\tilde {\mathbf \Phi}(s)
 \frac{1}{\mathbf I-\mathbf T \tilde {\mathbf \Psi}(s) }
\mathbf T \tilde {\mathbf \Psi}^{(0)}(s). \nonumber
\end{eqnarray}
This expression is quite general and valid for other models of $\mathbf T, 
\tilde {\mathbf \Phi}(s),  \tilde {\mathbf \Psi}(s)$, and  
$\tilde {\mathbf \Phi}^{(0)}(s),  \tilde {\mathbf \Psi}^{(0)}(s)$.
Further calculations are straightforward for the considered model, $T_{ij}=p_i$,  $\sum_i p_i=1$. 
The first term in (\ref{first})
contributes as $\langle x^2 \rangle_{\rm st}\Phi^{(0)}(t)$ to the correlation function. 
The series also can be summed
 exactly by using that $(\mathbf T \mathbf \Psi)_{ij}=p_i  \psi$, or 
$\mathbf T \mathbf \Psi=\psi \mathbf T$, and $(\mathbf T \mathbf \Psi)^2=\psi^2 \mathbf T$ repeatedly,
and standard relations between the densities and survival probabilities 
like $\tilde \Phi(s)=[1-\tilde \psi(s)]/s$.
The series term contributes as $\langle x\rangle^2_{\rm st} \tilde \psi^{(0)}(s)/s$, and
after some algebra we obtain the simple exact result
\begin{eqnarray}
\langle x(t)x(0)\rangle_{\rm st}=\langle \delta x^2 \rangle_{\rm st}
\Phi^{(0)}(t) + \langle x \rangle_{\rm st}^2,
\end{eqnarray}
with $\langle \delta x^2 \rangle_{\rm st}=\langle  x^2 \rangle_{\rm st}-\langle x \rangle_{\rm st}^2$.
This yields covariance (\ref{ACF}) upon identifying stationary averages 
with thermal equilibrium  averages. This result is valid for
any number of states within the fully decoupled CTRW model.

The simplest stationary process of this kind is two-state semi-Markov process
considered by Geisel \textit{et al.} \cite{Geisel} for velocity variable, as a statistical
model for chaotic dynamics. 
It must be stressed that $\psi(\tau)$ is not the
time probability density to stay in the corresponding state because at each scattering event
the particle can remain in the same state. This is waiting time distribution between 
scattering events. This process differs from the alternating symmetric two-state process
considered by Stratonovich, \textit{et al.} \cite{Stratonovich,GH03}. The last one alternates
at each scattering event with the probability one, i.e. $T_{11}=T_{22}=0$, $T_{12}=T_{21}=1$.
In such a case, $\psi(t)$ denoted for this model as $\psi_S(t)$ is indeed the time density to 
reside in each state.
Autocorrelation function of such a process 
looks less elegant:
\begin{eqnarray}
\tilde K(s)=\frac{\langle \delta x^2 \rangle_{\rm st}}{s}\left ( 
1-\frac{2}{\langle \tau_S\rangle s }\frac{1-\tilde \psi_S(s)}{1+\tilde \psi_S(s)} \right )\;.
\end{eqnarray}
This result can also be easily derived using the corresponding transition matrix $\mathbf T$.
It cannot be immediately inverted
to the time domain. However,
the relation between these two different two-state semi-Markovian models is in fact 
simple:
$\tilde \psi_S(s)=\tilde \psi(s)/[2-\tilde \psi(s)]$ with $\langle \tau_S\rangle =2\langle \tau\rangle $.

\end{document}